\documentclass{ws-procs9x6}

\newcommand{\collab}{(ATHENA Collaboration)}
\newcommand{\Hbar}{$\overline{\text{H}}$}
\newcommand{\pbar}{$\bar{p}$}
\newcommand{\pos}{$e^{+}$}
\newcommand{\el}{$e^{-}$}
\newcommand{\ket}[1]{|{#1}\rangle}

\begin{document}

\title{ATHENA -- First Production of Cold Antihydrogen and Beyond}

\author{A.~Kellerbauer$^{1}$, M.~Amoretti$^{2\mbox{,}3}$,
C.~Amsler$^{4}$, G.~Bonomi$^{1}$, P.~D.~Bowe$^{5}$,
C.~Canali$^{2\mbox{,}3}$, C.~Carraro$^{2\mbox{,}3}$, C.~L.~Cesar$^{6}$,
M.~Charlton$^{5}$, M.~Doser$^{1}$, A.~Fontana$^{7\mbox{,}8}$,
M.~C.~Fujiwara$^{9}$, R.~Funakoshi$^{10}$, P.~Genova$^{7\mbox{,}8}$,
J.~S.~Hangst$^{11}$, R.~S.~Hayano$^{10}$, I.~Johnson$^{4}$,
L.~V.~J{\O}rgensen$^{5}$, V.~Lagomarsino$^{2\mbox{,}3}$,
R.~Landua$^{1}$, E.~Lodi~Rizzini$^{12\mbox{,}8}$,
M.~Macr\'{i}$^{2\mbox{,}3}$, N.~Madsen$^{11}$,
G.~Manuzio$^{2\mbox{,}3}$, D.~Mitchard$^{5}$,
P.~Montagna$^{7\mbox{,}8}$, H.~Pruys$^{4}$, C.~Regenfus$^{4}$,
A.~Rotondi$^{7\mbox{,}8}$, G.~Testera$^{2\mbox{,}3}$, A.~Variola$^{5}$,
L.~Venturelli$^{12\mbox{,}8}$, D.~P.~van~der~Werf$^{5}$,
Y.~Yamazaki$^{9}$, and N.~Zurlo$^{12\mbox{,}8}$}

\address{$^{1}$Department of Physics, CERN, 1211~Gen\`{e}ve~23, Switzerland\\
$^{2}$Dipartimento di Fisica, Universit\`{a} di Genova, 16146~Genova,
Italy\\
$^{3}$INFN Sezione di Genova, 16146 Genova, Italy\\
$^{4}$Physik-Institut, University of Zurich, 8057 Z\"{u}rich,
Switzerland\\
$^{5}$Department of Physics, University of Wales Swansea, Swansea
SA2~8PP, UK\\
$^{6}$Instituto di Fisica, Universidade Federal do Rio de Janeiro, Rio
de Janeiro 21945-970, Brazil\\
$^{7}$Dipartimento di Fisica Nucleare e Teorica, Universit\`{a} di
Pavia, 27100 Pavia, Italy\\
$^{8}$INFN Sezione di Pavia, 27100 Pavia, Italy\\
$^{9}$Atomic Physics Laboratory, RIKEN, Saitama 351-0198, Japan\\
$^{10}$Department of Physics, University of Tokyo, Tokyo 113-0033,
Japan\\
$^{11}$Department of Physics and Astronomy, University of Aarhus,
8000~Aarhus~C, Denmark\\
$^{12}$Dipartimento di Chimica e Fisica per l'Ingegneria e per i
Materiali, Universit\`{a} di Brescia, 25123 Brescia, Italy}

\author{\protect\collab{}}

\maketitle

\abstracts{Atomic systems of antiparticles are the laboratories of
choice for tests of CPT symmetry with antimatter. The ATHENA experiment
was the first to report the production of copious amounts of cold
antihydrogen in 2002. This article reviews some of the insights that
have since been gained concerning the antihydrogen production process
as well as the external and internal properties of the produced
anti-atoms. Furthermore, the implications of those results on future
prospects of symmetry tests with antimatter are discussed.}

\section{Introduction}

According to the CPT theorem\cite{bib:paul1957}, all physical laws are
invariant under the combined operations of charge conjugation, parity
(reversal of the spatial configuration), and time reversal. Since CPT
transforms an elementary particle into its antiparticle, their
fundamental properties such as mass, charge, and magnetic moment, are
either exactly equal or exactly opposed. This predestines antimatter
for tests of CPT symmetry. Due to the fact that atomic spectroscopy on
the transition between the ground and first excited states (1s--2s) of
hydrogen has been carried out to $10^{-14}$ relative
precision\cite{bib:nier2000}, this transition is also being targeted
for CPT tests with hydrogen and antihydrogen (\Hbar{}). In addition to
atomic spectroscopy, antimatter gravity tests are being considered.
These could be carried out by gravity interferometry\cite{bib:phil1997}
or in atomic fountains, in analogy with current studies on ordinary
matter in such setups\cite{bib:pete1999}.

Two dedicated experiments, ATHENA\cite{bib:amor2004a} and
ATRAP\cite{bib:gabr1999b}, have been set up at CERN's Antiproton
Decelerator (AD)\cite{bib:heme1999} since 1998 with the goal of
producing sufficient amounts of antihydrogen to ultimately allow
precision atomic spectroscopy and a comparison of its atomic spectrum
with that of hydrogen. A production of large amounts of \Hbar{} was
first demonstrated by ATHENA\cite{bib:amor2002} and later by
ATRAP\cite{bib:gabr2002}, using very similar schemes for antihydrogen
production but different detection techniques. In the time since these
independent proofs of principle, the main challenges have been to
investigate the parameters that govern efficient \Hbar{} production and
the internal and external properties of the produced antihydrogen.

\section{Setup and principle}

The ATHENA apparatus\cite{bib:amor2004a} consists of three main
components, shown in Fig.~\ref{fig:Overview}:
\begin{figure}
  \centering
  \includegraphics{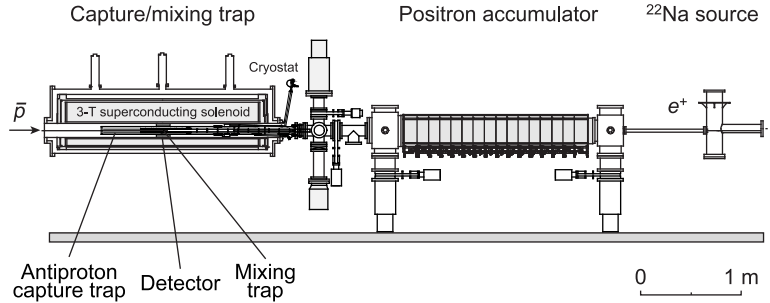}%
  \caption{Overview of the ATHENA apparatus. Shown on the left is the
superconducting 3-T solenoid magnet which houses the capture trap, the
mixing trap, and the antihydrogen annihilation detector. On the right,
the radioactive sodium source for the positron production and
the 0.14-T positron accumulation Penning trap.%
\label{fig:Overview}}
\end{figure}
The antiproton (\pbar{}) capture trap, the mixing trap, and the
positron (\pos{}) source and accumulator. The former two are located in
the 3-T field of a superconducting magnet whose bore is kept at
liquid-nitrogen temperature. A liquid-helium cryostat, whose cold nose
protrudes into the magnet bore and encloses the trap, reduces the
temperature of the trap region further to about 15~K.

The bunch of about $2\mbox{--}3 \times 10^{7}$ antiprotons that is
extracted from the AD after every deceleration and cooling cycle
undergoes a final deceleration step in a thin ($\approx 50$~$\mu$m)
degrader foil. The foil's thickness is chosen in order to optimize the
fraction of \pbar{} that can be trapped by the capture trap's
high-voltage electrodes (5~kV potential). In the capture trap, the
confined antiprotons cool in Coulomb collisions with an electron plasma
that was loaded prior to the \pbar{} capture and allowed to cool by
emission of synchrotron radiation. Typically, two \pbar{} spills from
the AD are stacked in the capture trap, resulting in about~$10^{4}$
\pbar{} ready for mixing. Simultaneously, positrons produced in the
$\beta$~decay of the radionuclide $^{22}$Na are moderated, then cooled
in collisions with nitrogen buffer gas and accumulated in a low-field
Penning trap at room temperature.

\begin{figure}
  \centering
  \includegraphics{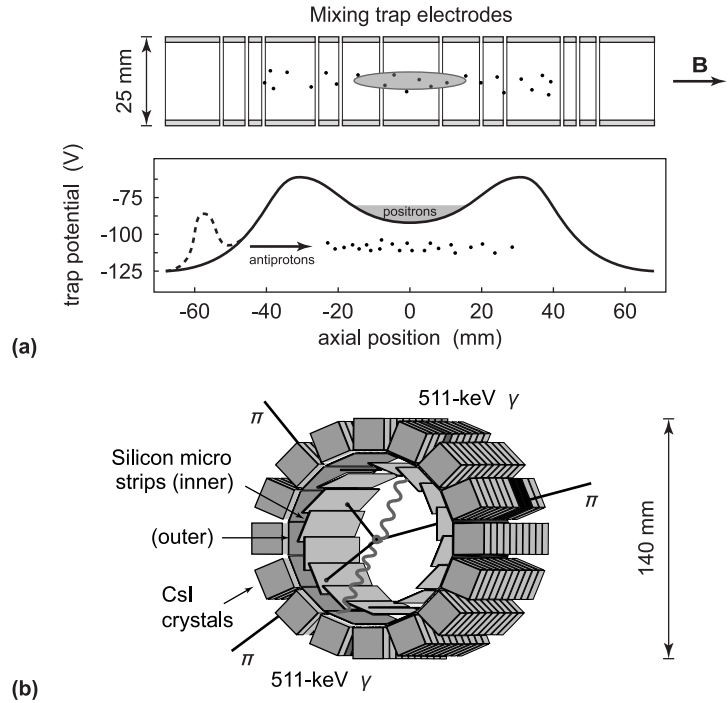}%
  \caption{(a)~Detailed sketch of the mixing trap, which is operated
in a nested-trap configuration. The graph shows the axial trap
potential before (dashed line) and after (solid line) the antiproton
injection. (b)~Sketch of the antihydrogen annihilation detector. With
its highly granular silicon strip and CsI crystal modules, it allows a
direct and unambiguous detection of
\Hbar{} production.%
\label{fig:Trap_and_detector}}
\end{figure}

After the independent \pbar{} stacking and \pos{} accumulation phases,
the axial potential in the mixing trap is brought into a so-called
nested configuration\cite{bib:gabr1988}.
Figure~\ref{fig:Trap_and_detector}(a)
shows how this potential shape allows both positively and negatively
charged particles to be simultaneously confined. The central well of
this nested trap is then first filled with the $\approx 5 \times
10^{7}$ accumulated positrons. Just like the electrons in the capture
trap, these cool to the ambient temperature of about 15~K by emitting
synchrotron radiation. The antiprotons are transferred into a small
lateral well and then launched into the mixing region with a relative
energy of 30~eV. They oscillate between the lateral confines of the
nested well, repeatedly traverse the \pos{} plasma, and rapidly cool in
Coulomb collisions with the positrons. After some tens of~ms,
antihydrogen production spontaneously sets in with initial rates of
several~100~Hz.

The neutral \Hbar{} atoms thus produced in the center of the mixing
trap are no longer affected, to first order, by the electrical and
magnetic fields used for the charged-particle confinement. They leave
the interaction region with a momentum that is essentially equal to
that of the \pbar{} just before \Hbar{} formation. When these
anti-atoms impinge upon the Penning trap electrodes, their constituents
immediately annihilate with ordinary matter. The signal of these
destructive events is recorded with the antihydrogen annihilation
detector that surrounds the mixing trap. A sketch of this detector is
shown in Fig.~\ref{fig:Trap_and_detector}(b). It consists of 8192
silicon strips in two layers for the detection of the charged pions
created in the \pbar{} annihilation with a proton or a neutron and 192
cesium iodide crystals that record the (mainly back-to-back)
$\gamma$~rays from the \pos{}--\el{} annihilation. Despite its
extremely compact dimensions, it allows a three-dimensional
reconstruction of the charged-particle vertex with a resolution of 4~mm
and a spatial and temporal correlation of the \pbar{} and \pos{}
signals for an unambiguous identification of \Hbar{} formation.

\begin{figure}
  \centering
  \includegraphics{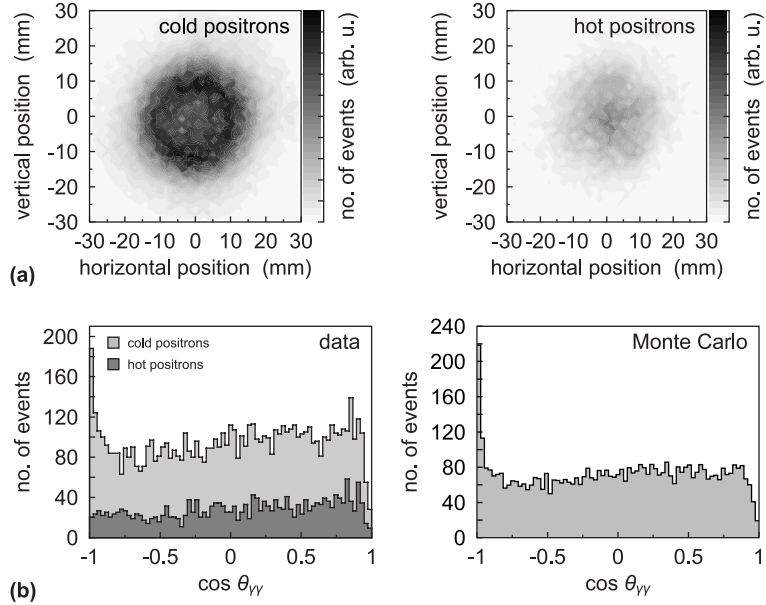}%
  \caption{Signal of the first production of cold antihydrogen with
ATHENA\protect\cite{bib:amor2002}. (a)~Charged-pion vertex distribution
as a function of the azimuthal coordinates. (b)~Opening-angle
distribution of the photons recorded in coincidence with the
charged-particle hits, as seen from the charged-particle vertex.%
  \label{fig:Detector_signal}}%
\end{figure}

As an example for event reconstruction, Fig.~\ref{fig:Detector_signal}
illustrates the signal of the first production of cold
antihydrogen\cite{bib:amor2002} in 2002. In
Fig.~\ref{fig:Detector_signal}(a), the azimuthal distribution of
reconstructed vertices from the \pbar{} annihilation is shown. In the
left panel, where the positrons are in thermal equilibrium with the
trap at 15~K (``cold mixing''), the largest numbers of events are
recorded in a ring located at the position of the electrodes (25~mm
diameter). When the positron plasma is heated by means of a
radiofrequency (RF) excitation of the axial plasma
modes\cite{bib:amor2003} (``hot mixing''), \Hbar{} production is
suppressed and a smaller number of events, due to \pbar{} annihilations
with residual gas, is recorded in the center of the trap (right panel).
Figure~\ref{fig:Detector_signal}(b) shows the distribution of the
opening angle of the two 511-keV $\gamma$~rays recorded in time
coincidence with the charged-particle hits, as seen from the
charged-particle vertex. In the left panel, a clear excess at an
opening angle of 180$^{\circ}$ is present for cold positrons, while it
is suppressed when the positron plasma is heated. The right panel shows
the good agreement of the data with Monte-Carlo simulations. The
131(22) fully reconstructed events that constitute the peak in the left
panel of Fig.~\ref{fig:Detector_signal}(b) correspond to a total number
of about~50\hspace{0.25em}000 produced \Hbar{} atoms for this partial
dataset of 2002. A complete analysis of the 2002 data, together with
more detailed Monte Carlo simulations, showed that the instantaneous
trigger rate from the silicon detector is a good proxy for antihydrogen
production, with 65\% of all triggers over the entire mixing cycle due
to annihilating antihydrogen atoms\cite{bib:amor2004b}.

\section{Recent results}

For precise antimatter studies, it is not sufficient to merely produce
large numbers of antihydrogen. A fair knowledge of the temperature and
kinetic-energy distributions of the produced \Hbar{} is required in
order to estimate the fraction of anti-atoms which can be trapped. The
\Hbar{} atoms must also be produced in a well-defined internal quantum
state, if possible the ground state. In this section, some results on
the latter question, based on an investigation of the \Hbar{} formation
process, are presented.

\subsection{Antihydrogen production 2002/2003}

As a prerequisite for any quantitative studies on antihydrogen
formation, the offline data analysis must allow a precise determination
of the number of produced \Hbar{} anti-atoms. In order to achieve this,
one or several observables, such as the radial vertex distribution or
the 2$\gamma$ opening angle distribution, can be considered as a linear
combination of a pure \Hbar{} signal (Monte-Carlo simulation of
annihilations on the trap electrodes) and background. Since the
background is expected to be mainly due to \pbar{} annihilations with
residual gas, it can be represented by the signal obtained from runs in
which the \pos{} were heated to several 1000~K, thereby inhibiting
\Hbar{} production. The total antihydrogen production of 2002 and 2003
obtained in this way is summarized in Tab.~\ref{tab:production}.
\begin{table}
  \tbl{Comparative summary of ATHENA's antihydrogen production in 2002
       and 2003.%
  \label{tab:production}}%
{%
    \begin{tabular}{lll}
\hline
\rule{0pt}{3.5ex} ~ & \multicolumn{1}{c}{Cold mixing 2002} & \multicolumn{1}{c}{Cold mixing 2003}\\[1ex]
\hline
\rule{0pt}{3.5ex}Total no.\ of cycles & 341 & 416\\[1ex]
Cycle duration & 180~s & 70~s\\[1ex]
Total mixing time & 17.1~h & 8.1~h\\[1ex]
Injected \pbar{} & $2.92 \times 10^{6}$ & $5.07 \times 10^{6}$\\[1ex]
Produced \Hbar{} & $4.94 \times 10^{5}$ & $7.04 \times 10^{5}$\\[1ex]
Production efficiency &  16.9\% & 13.9\%\\[1ex]
Avg.\ \Hbar{} production rate & 8.0(4)~Hz & 24.2(1.3)~Hz\\[1ex]
\Hbar{} fraction of signal& 65(5)\% & 74(3)\%\\[1ex]
\hline
    \end{tabular}%
}
\end{table}
It shows that ATHENA has produced more than~$10^{6}$ \Hbar{} anti-atoms
since its start of operations and that the production efficiency in
terms of captured antiprotons from the AD is between 10~and 20\%.

\subsection{Recombination process}

The formation of antihydrogen by direct capture of a positron onto an
atomic orbit around an antiproton does not simultaneously conserve
energy and momentum. The involvement of a third particle is needed in
order to respect these conservation laws. That particle can either be a
photon in the case of (spontaneous) radiative recombination
(SRR)\cite{bib:stev1975} or a second positron in three-body
recombination (TBR)\cite{bib:glin1991}. These two processes are
predicted to have vastly different cross-sections and recombination
rates, with TBR expected to be the dominant process at ATHENA's
experimental conditions. The most important difference with a view to
precision studies lies in the fact that SRR populates low-lying states
($n < 10$) and TBR highly excited Rydberg states ($n \gg 10$). The two
mechanisms also exhibit different dependencies on the positron
temperature (SRR: $\propto T^{-0.63}$; TBR: $\propto T^{-4.5}$) and
density (SRR: $\propto n_{e^{+}}$; TBR: $\propto n^{2}_{e^{+}}$), which
can allow to distinguish between them.

\begin{figure}
  \centering
  \includegraphics{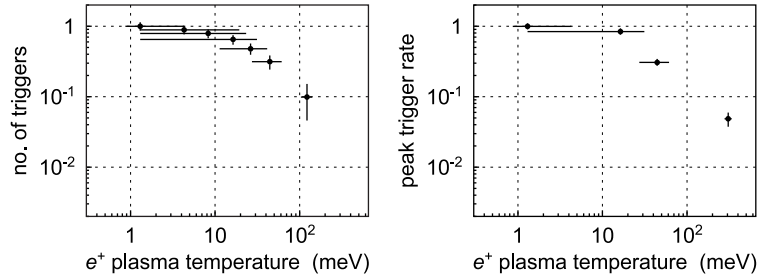}%
  \caption{Dependence of the background-corrected integrated total\
number of charged-particle triggers per mixing cycle (left) and the
peak trigger rate (left) on the positron plasma
temperature\protect\cite{bib:amor2004d}. The number of triggers and
trigger rate have been normalized to the signal for an \pos{}
temperature of 1~meV. Note the logarithmic scale.%
\label{fig:temp_dependence}}
\end{figure}

In order to determine the temperature dependence of \Hbar{} production,
we have performed mixing cycles with RF~heating at various amplitudes
applied to the positron plasma\cite{bib:amor2004d}. The positron
temperature increase was measured with ATHENA's plasma diagnostics
system\cite{bib:amor2003} by resonant excitation and detection of the
axial plasma modes. In Fig.~\ref{fig:temp_dependence}, the
background-corrected integrated number of triggers (left) and peak
trigger rate (right) as possible proxies for \Hbar{} production are
shown as a function of the positron temperature, assuming an
equilibrium temperature of 15~K.
Neither of these plots shows the characteristics of a simple power law
(a straight line in these logarithmic plots), but a best fit to the
data yields a behavior of the form $\propto T^{-0.7(2)}$, close to that
expected from radiative recombination. However, the observed event
rates are between~1 and~2 orders of magnitude higher than expected for
this recombination process.

The second access to the recombination process is via the positron
density dependence. For this purpose, we have analyzed the 2003 data
with a view to varying positron plasma density\cite{bib:bono2005}.
Standard cold mixing runs with positron plasma densities between $3
\times 10^{8}$/cm$^{3}$ and $1.5 \times 10^{9}$/cm$^{3}$ have been
identified. However, this analysis is complicated by the fact that
under ATHENA's typical experimental conditions, the \pbar{} cloud has a
much larger radial extent than the \pos{} plasma. This means that the
number of interacting antiprotons strongly depends on their radial
density distribution. Measurements of this distribution are therefore
required to extract the positron density dependence of \Hbar{}
production.

\section{Antihydrogen spectroscopy within the framework of the
Standard-Model Extension}

Any measured difference in the hydrogen and antihydrogen atomic spectra
would be a clear and unambiguous signal for CPT violation. On the other
hand, a theoretical framework for such symmetry breaking can indicate
which transitions are particularly suited for an experimental search.
The Standard Model Extension\cite{bib:coll1997} incorporates
spontaneous CPT and Lorentz breaking at a fundamental level. It is an
extension of the Standard Model that preserves energy and momentum
conservation, gauge invariance, renormalizability, and micro-causality.

\begin{figure}
  \centering
  \includegraphics{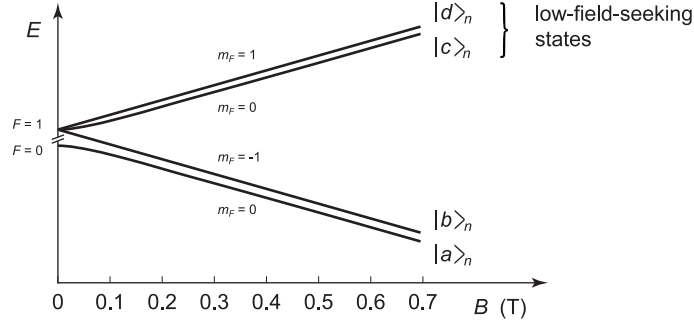}%
  \caption{Hyperfine and
Zeeman splitting of (anti)hydrogen confined in a magnetic field for
states with zero angular momentum.%
\label{fig:zeeman}}
\end{figure}

Within this framework, the sensitivity to CPT- and Lorentz-violating
terms of spectroscopic experiments on H and \Hbar{} confined in a
magnetic trap can be predicted. Consider the energy states of
(anti)hydrogen with zero angular momentum ($l = 0$), confined in a
magnetic trap with axial solenoidal field and radial multipole magnetic
fields. These states are subject to hyperfine as well as Zeeman
splitting, as shown in Fig.~\ref{fig:zeeman}.
Before the excitation, only the low-field-seeking states $\ket{c}_{1}$
and $\ket{d}_{1}$ are confined in a magnetic trap. It has been
shown\cite{bib:bluh1999} that the CPT-violating term $b^{e}_{3}$ shifts
all $\ket{d}_{n}$ states by the same amount, both in~H and in~\Hbar{},
and the CPT-violating effect is thus suppressed by a factor $\alpha^{2}
/ 8 \pi$ in the $\ket{d}_{1} \longrightarrow \ket{d}_{2}$ transition.
The transition $\ket{c}_{1} \longrightarrow \ket{c}_{2}$ between the
mixed-spin states does potentially produce an unsuppressed frequency
shift due to the $n$~dependence of the hyperfine splitting. This shift
is different in~H and in~\Hbar{} and leads both to diurnal variations
in the frequency difference and to a non-zero instantaneous difference.
However, this transition is field-dependent and thus subject to Zeeman
broadening in inhomogeneous magnetic fields.

As an alternative, it was therefore suggested\cite{bib:bluh1999} to
consider a transition between hyperfine levels of the ground state ($n
= 1$) at the field-independent transition point $B \approx 0.65$~T. The
transition $\ket{d}_{1} \longrightarrow \ket{c}_{1}$ in~H and
in~\Hbar{} is then subject to potential diurnal variations, and the
instantaneous difference $\Delta \nu_{cd}$ between these transitions in
hydrogen and antihydrogen is directly proportional to the CPT-violating
term $b^{p}_{3}$. Based on these considerations, future antihydrogen
spectroscopy experiments will include comparisons of the hydrogen and
antihydrogen hyperfine structure\cite{bib:haya2005}.

\section{Conclusions and outlook}

With the first production of copious amounts of cold antihydrogen, many
of the challenges on the way to high-precision CPT tests with
antimatter have been surmounted, but many more still remain. Future
high-precision spectroscopic and interferometric measurements on
antimatter atoms are contingent upon the ability to confine neutral
\Hbar{} atoms and to cool them with Lyman-$\alpha$ lasers. Our results
on the temperature dependence of \Hbar{} production suggest on the one
hand that an appreciable fraction of the antihydrogen may be produced
in low-lying states accessible to precision atomic spectroscopy. On the
other hand, recombination possibly sets in before complete
thermalization of the antiprotons, thereby reducing the fraction of
produced antihydrogen that can be confined in a magnetic trap. Further
studies on antihydrogen production in a nested Penning trap are
required to clarify these points. In parallel, tests with ordinary
matter on the simultaneous confinement of charged and neutral particles
in electromagnetic traps are being carried out in order to establish
parameters for the efficient preparation of trapped antihydrogen for
symmetry tests.

\section*{Acknowledgments}

This work was supported by the funding agencies INFN (Italy), CNPq
(Brazil), MEXT (Japan), SNF (Switzerland), SNF (Denmark), and EPSRC
(United Kingdom).

\end{document}